\documentclass[a4paper,11pt,twoside]{article}
\usepackage{graphicx,nfg}

\begin{document}
\makeatletter
\renewcommand{\theequation}{\thesection.\arabic{equation}}
\@addtoreset{equation}{section}
\makeatother

\title{\sc\huge ${\cal N}$-fold Supersymmetry\\ in
Quantum Mechanics\\ - General Formalism -}
\author{\Large Hideaki
Aoyama$^{\dagger 1}$,
Masatoshi Sato$^{\ddagger 2}$
and
Toshiaki Tanaka$^{\dagger 3}$
, \\[8pt]
\sc $^\dagger$Faculty of Integrated Human Studies\\[4pt]
\sc Kyoto University, Kyoto 606-8501, Japan\\[7pt]
\sc $^\ddagger$The Institute for Solid State Physics\\[4pt]
\sc The University of Tokyo, Kashiwanoha 5-1-5, \\[4pt]
\sc Kashiwa-shi, Chiba 277-8581, Japan}
\footnotetext[1]{aoyama@phys.h.kyoto-u.ac.jp}
\footnotetext[2]{msato@issp.u-tokyo.ac.jp}
\footnotetext[3]{ttanaka@phys.h.kyoto-u.ac.jp}
\maketitle

\vspace{-10cm}
\rightline{KUCP-0182}
\vspace{9.5cm}
\thispagestyle{empty}
\def\nfsusy{${\cal N}$-fold supersymmetry}
\begin{abstract}
We report general properties of ${\cal N}$-fold supersymmetry in
one-dimensional quantum mechanics.
${\cal N}$-fold supersymmetry is characterized by supercharges
which are ${\cal N}$-th polynomials of momentum.
Relations between the anti-commutator of the supercharges and the
Hamiltonian,  the spectra,  the Witten index, the non-renormalization
theorems and the quasi-solvability are examined.   
We also present further investigation about a particular class of 
${\cal N}$-fold supersymmetric models which we dubbed type A.
Algebraic equations which determine a part of spectra of type A models
are presented, and the non-renormalization theorem are generalized.
Finally, we present a possible generalization of ${\cal N}$-fold
 supersymmetry in multi-dimensional quantum mechanics. 
\end{abstract}
\newpage

\section{Introduction}
One of unique aspects of supersymmetry is its usefulness as a tool of
non-perturbative analyses of quantum theories. 
Various non-renormalization theorems enable us to reveal
non-perturbative properties of quantum theories without annoyance of
perturbative corrections. 
An approach to quark confinement problem via ${\cal N}=2$ supersymmetric QCD
\cite{SW} is a good example which represents this aspect.
In Refs.\cite{AKOSW,AKOSW2}, a method of calculation of
non-perturbative part of the energy spectrum was developed and tested
with aid of supersymmetry. It is based on the valley method
\cite{RR}-\cite{AKHOSW}, and together with an understanding of the
Bogomolny technique \cite{Bog}, it correctly led to an explanation of 
the disappearance of
the leading Borel singularity of the perturbative corrections for the
ground energy when the theory becomes supersymmetric: 
Since the ground state of the supersymmetric theories does not receive any
perturbative corrections \cite{Wit,Wit2}, the Borel singularity must
vanish in this case. 

The method also predicted the disappearance
of the leading Borel singularity of the perturbative corrections 
at other values of a parameter in the theory, which do not
correspond to the case when the theory becomes supersymmetric. 
This disappearance of the leading Borel singularity was understood by an
extension of supersymmetry, which was named  ``${\cal N}$-fold
supersymmetry'' \cite{AKOSW2}, which  
supercharges are ${\cal N}$-th
polynomials of momentum.
When ${\cal N}=1$, they reduce to ordinary supersymmetry.
Similar higher derivative generalizations of supercharges were
investigated in various different contexts \cite{AIS}-\cite{FNN}. 

In this paper, we investigate general properties of ${\cal N}$-fold
supersymmetry. 
First in section \ref{dons}, we define ${\cal N}$-fold
supersymmetry in one-dimensional quantum mechanics and fix notations
used throughout in this paper. 

In section \ref{sec_gp_mh}, we introduce the ``Mother Hamiltonian'' as
the anti-commutator of the supercharges.
In contrast to ordinary supersymmetry, it does not coincide with the
Hamiltonian in general.    
Relations between the ordinary Hamiltonian and the Mother Hamiltonian 
are shown.  
Spectra of the ${\cal N}$-fold supersymmetric
systems are examined in section \ref{sec_gp_s}.
We investigate a relation between ${\cal N}$-fold supersymmetry and
polynomial supersymmetry \cite{AIN} in section
\ref{sec_p_ps}.
The Witten index is generalized to ${\cal
N}$-fold supersymmetry in section \ref{sec_gp_gwi}.
In section \ref{sec_gp_nt}, non-renormalization theorems for ${\cal
N}$-fold supersymmetry are briefly discussed.
For ${\cal N}$-fold supersymmetric systems,  non-renormalization
theorems hold as well as ordinary supersymmetric ones.
In section \ref{sec_gp_qans}, we show a close relation between
quasi-solvability and  ${\cal N}$-fold supersymmetry. 
For ${\cal N}$-fold supersymmetric systems, a part of spectra (not
complete spectra) can be solvable.
We show that quasi-solvability is equivalent to ${\cal N}$-fold
supersymmetry. 

Two examples of ${\cal N}$-fold supersymmetric systems are illustrated
in section \ref{e}. 
As the simplest but non-trivial example of ${\cal N}$-fold
supersymmetry, 2-fold supersymmetry are examined in section \ref{sec_e_2s}.
In section \ref{sec_e_typea}, a class of ${\cal N}$-fold supersymmetric
systems which we dubbed ``type A'' \cite{AST} is investigated. 
The type A models include ${\cal N}$-fold supersymmetric systems found
in Refs.\cite{AKOSW2, ASTY}.
For the type A models, a part of the spectra are determined by algebraic
equations. 
Using this equations, the non-renormalization theorem found in
Refs.\cite{AKOSW2, ASTY} are generalized to most of type A models.

In section \ref{nsims}, we suggest a possible generalization of
${\cal N}$-fold supersymmetry in multi-dimensional systems.

Finally, in appendix \ref{appendixa} we examine the coupling constant
dependence in the type A models.

\section{Definition of ${\cal N}$-fold supersymmetry}
\label{dons}
Let us first define ${\cal N}$-fold
supersymmetry in one-dimensional quantum mechanics.   
To define the ${\cal N}$-fold supersymmetry, we introduce the
following Hamiltonian ${\bf H}_{\cal N}$,
\begin{eqnarray}
{\bf H}_{\cal N}=H^{-}_{{\cal N}}\psi\psi^{\dagger}
+H^{+}_{{\cal N}}\psi^{\dagger}\psi, 
\end{eqnarray}
where $\psi$ and $\psi^{\dagger}$ are fermionic coordinates which satisfy
\begin{eqnarray}
\{\psi,\psi\}=\{\psi^{\dagger}, \psi^{\dagger}\}=0
,\quad 
\{\psi,\psi^{\dagger}\}=1,
\end{eqnarray}
and $H^{\pm}_{\cal N}$ are ordinary Hamiltonians, 
\begin{eqnarray}
H^{-}_{\cal N}=\frac{1}{2}p^{2}+V^{-}_{\cal N}(q)
,\quad
H^{+}_{\cal N}=\frac{1}{2}p^{2}+V^{+}_{\cal N}(q),
\end{eqnarray}
where $p=-id/dq$.
The ${\cal N}$-fold supercharges are generically defined as 
\begin{eqnarray}
Q_{\cal N}=P^{\dagger}_{\cal N}\psi
,\quad
Q^{\dagger}_{\cal N}=P_{\cal N}\psi^{\dagger}, 
\end{eqnarray}
where $P_{\cal N}$ is an ${\cal N}$-th order polynomial of $p$,
\begin{eqnarray}
P_{\cal N}=
w_{\cal N}(q)p^{\cal N}+w_{{\cal N}-1}(q)p^{{\cal N}-1}+\cdots+w_1(q)p+w_0(q). 
\label{sec_dons:pn}
\end{eqnarray}
A system is defined to be ${\cal N}$-fold supersymmetric if the following
${\cal N}$-fold supersymmetric algebra is satisfied,
\begin{eqnarray}
&&\{Q_{\cal N}, Q_{\cal N}\}=\{Q^{\dagger}_{\cal N}, Q^{\dagger}_{\cal N}\}=0,
\\
&&[Q_{\cal N}, {\bf H}_{\cal N}]=[Q^{\dagger}_{\cal N}, {\bf H}_{\cal N}]=0. 
\end{eqnarray}
The former relation is trivially satisfied, but the latter gives the
following conditions, 
\begin{eqnarray}
P_{\cal N}H^{-}_{\cal N}-H^{+}_{\cal N}P_{\cal N}=0
,\quad
P^{\dagger}_{\cal N}H^{+}_{\cal N}-H^{-}_{\cal N}P^{\dagger}_{\cal N}=0.
\label{sec_dons:n-fold_alg2}
\end{eqnarray}
These conditions generally give ${\cal N}+2$ differential equations for 
${\cal N}+3$ functions $V^{-}_{\cal N}(q)$,
$V^{+}_{\cal N}(q)$ and $w_{n}(q)\,(n=0,\cdots, {\cal N})$, thus 
one function remains arbitrary.
We obtain the equation $w_{\cal N}'(q)=0$ 
by comparison of the coefficient of the $\partial^{{\cal N}+1}$ terms in
Eq.(\ref{sec_dons:n-fold_alg2}). 
Thus we can set $w_{\cal N}(q)=1$ without losing generality.

The above definition of ${\cal N}$-fold supersymmetry includes
ordinary supersymmetry \cite{Wit, Wit2}, which is realized when ${\cal N}=1$, 
$w_{0}(q)=-iW(q)$ and 
\begin{eqnarray}
V^{-}_{\cal N}(q)=\frac{1}{2}\left(W(q)^2-W'(q)\right)
,\quad
V^{+}_{\cal N}(q)= \frac{1}{2}\left(W(q)^2+W'(q)\right).
\end{eqnarray} 

Conveniently, $\psi$ 
and $\psi^{\dagger}$ are often represented as the following $2\times 2$
matrix form, 
\begin{eqnarray}
\psi=
\left(
\begin{array}{cc}
0 & 0 \\
1 & 0
\end{array}
\right)
,\quad
\psi^{\dagger}
= 
\left(
\begin{array}{cc}
0 & 1 \\
0 & 0
\end{array}
\right).
\label{sec_dons_matrix1}
\end{eqnarray}
In this notation, the ${\cal N}$-fold supercharges are given by
\begin{eqnarray}
Q_{\cal N}=\left(
\begin{array}{cc}
0&0 \\
P^{\dagger}_{\cal N}&0
\end{array} 
\right)
,\quad
Q^{\dagger}_{\cal N}=\left(
\begin{array}{cc}
0&P_{\cal N} \\
0&0
\end{array} 
\right),
\label{sec_dons_matrix2}
\end{eqnarray}
and the Hamiltonian is given by
\begin{eqnarray}
{\bf H}_{\cal N}=
\left(
\begin{array}{cc}
H^{+}_{\cal N} & 0\\
0 & H^{-}_{\cal N}
\end{array}
\right). 
\label{sec_dons_matrix3}
\end{eqnarray}
We define the fermion number operator $F$ as follows,
\begin{eqnarray}
F=\psi^{\dagger}\psi=
\left(
\begin{array}{cc}
1 &0 \\
0& 0
\end{array} 
\right).
\label{sec_dons_matrix4}
\end{eqnarray}
Thus the form of fermionic states is  
\begin{eqnarray}
\left(
\begin{array}{c}
\Phi^{+}\\
0
\end{array}
\right), 
\end{eqnarray}
and that of bosonic ones is
\begin{eqnarray}
\left(
\begin{array}{c}
0\\
\Phi^{-}
\end{array}
\right).
\end{eqnarray}
$H^{-}_{\cal N}$ and $H^{+}_{\cal N}$ are therefore the Hamiltonians of
bosonic states and fermionic ones respectively. 
For physical states, $\Phi^{\pm}$ are normalizable (square integrable)
functions on ${\bf R}$ or on a subset of ${\bf R}$.

\section{General properties}
\label{gp}

\subsection{Mother Hamiltonian}
\label{sec_gp_mh}

In systems with ordinary supersymmetry, the Hamiltonian is given by  
the anti-commutator of the supercharges. 
In systems with ${\cal N}$-fold supersymmetry, however, this
relation does not hold in general.
This is evident from the fact that $Q_{\cal N}$ contains 
${\cal N}$-derivatives with respect to the coordinate $q$ and therefore 
$\frac{1}{2}\{Q_{\cal N}^{\dagger}, Q_{\cal N}\}$ contains 
$2{\cal N}$-derivatives. 
In ${\cal N}$-fold supersymmetric systems, the anti-commutator has a
``family resemblance'' to the Hamiltonian, and is thus called 
``{\em Mother Hamiltonian}'': 
\begin{eqnarray}
{\cal H}_{\cal N}=\frac{1}{2}\{Q_{\cal N}^{\dagger}, Q_{\cal N}\} 
=\frac{1}{2}\left(
\begin{array}{cc}
P_{\cal N}P^{\dagger}_{\cal N} &0 \\
0 & P^{\dagger}_{\cal N} P_{\cal N}
\end{array}
\right).
\end{eqnarray}
The Mother Hamiltonian commutes with the ${\cal N}$-fold
supercharges, 
\begin{eqnarray}
[{\cal H}_{\cal N}, Q_{\cal N}]
= [{\cal H}_{\cal N}, Q^{\dagger}_{\cal N}]
=0.
\end{eqnarray}
To examine relations between the Mother Hamiltonian and the original
one,  
let us introduce ${\cal N}$ linearly independent functions $\phi^{-}_{n}(q)$
$(n=1,\cdots {\cal N})$ which satisfy the following relation,   
\begin{eqnarray}
P_{\cal N}\phi^{-}_{n}=0.
\end{eqnarray}
From Eq.(\ref{sec_dons:n-fold_alg2}), the following relation holds,
\begin{eqnarray}
P_{\cal N}H^{-}_{\cal N}\phi^{-}_{n}=0, 
\end{eqnarray}
thus $H^{-}_{\cal N}\phi^{-}_{n}$ is given by a linear
combination of $\phi^{-}_{n}$.  
We can therefore define the matrix ${\bf S}^{-}$ as follows,
\begin{eqnarray}
H^{-}_{\cal N}\phi^{-}_{n}=\sum_{m}{\bf S}^{-}_{n,m}\phi^{-}_{m}.
\end{eqnarray}
In a similar manner, for ${\cal N}$ independent functions which satisfy
\begin{eqnarray}
P^{\dagger}_{\cal N}\phi^{+}_{n}=0, 
\end{eqnarray}
the next equation holds,
\begin{eqnarray}
P^{\dagger}_{\cal N}H^{+}_{\cal N}\phi^{+}_{n}=0.
\end{eqnarray}
Thus we define ${\bf S}^{+}$ as follows,
\begin{eqnarray}
H^{+}_{\cal N}\phi^{+}_{n}=\sum_{m}{\bf S}^{+}_{n,m}\phi^{+}_{m}. 
\end{eqnarray}

From these matrices ${\bf S}^{\pm}$, the Mother Hamiltonian
${\cal H}_{\cal N}$ is given as follows,
\begin{eqnarray}
{\cal H}_{\cal N}=
\frac{1}{2}
{\rm det}{\bf M}_{\cal N}^{-}(H_{\cal N}^{-})\psi\psi^{\dagger}
+\frac{1}{2}
{\rm det}{\bf M}_{\cal N}^{+}(H_{\cal N}^{+})\psi^{\dagger}\psi 
+\frac{1}{2}
q^{+\dagger}Q_{\cal N}
+\frac{1}{2}
q^{-} Q_{\cal N}^{\dagger},
\end{eqnarray}
where
\begin{eqnarray}
\frac{1}{2}{\bf M}^{-}_{\cal N}(E)\equiv E{\bf I}-{\bf S}^{-}
,\quad 
\frac{1}{2}{\bf M}^{+}_{\cal N}(E)\equiv E{\bf I}-{\bf S}^{+},
\label{sec_p:misnit}
\end{eqnarray}
and $q^{\pm}$ are supercharges for at most (${\cal N}-1$)-fold
supersymmetry.
In 2$\times$2 matrix notation as in
Eqs.(\ref{sec_dons_matrix1})-(\ref{sec_dons_matrix4}), this becomes
\begin{eqnarray}
{\cal H}_{\cal N}=\frac{1}{2}
\left(
\begin{array}{cc}
{\rm det}{\bf M}^{+}_{\cal N}(H^{+}_{\cal N})+p^{+}P_{\cal N}^{\dagger}  & 0\\
0 & {\rm det}{\bf M}^{-}_{\cal N}(H^{-}_{\cal N})+p^{-\dagger}P_{\cal N} 
\end{array}
\right),
\label{sec_p_mh:mh_oh}
\end{eqnarray}
where $p^{\pm}$ are defined by
\begin{eqnarray}
q^{\pm}=
\left(
\begin{array}{cc}
0 &0 \\
p^{\pm\dagger} & 0
\end{array}
\right).
\end{eqnarray}
And if the ${\cal N}$-fold supercharges are uniquely determined for given
$H^{\pm}_{\cal N}$, the Mother Hamiltonian  
${\cal H}_{\cal N}$ has the following more simple form, 
\begin{eqnarray}
{\cal H}_{\cal N}
&=&\frac{1}{2}
{\rm det}{\bf M}^{-}_{\cal N}(H^{-}_{\cal N})\psi\psi^{\dagger}  
+\frac{1}{2}
{\rm det}{\bf M}^{+}_{\cal N}(H^{+}_{\cal N})\psi^{\dagger}\psi
\nonumber\\
&=&\frac{1}{2}
\left(
\begin{array}{cc}
{\rm det}{\bf M}^{+}_{\cal N}(H^{+}_{\cal N})  & 0\\
0 & {\rm det}{\bf M}^{-}_{\cal N}(H^{-}_{\cal N}) 
\end{array}
\right).
\label{sec_p_mh:poly}
\end{eqnarray}

\noindent
\underline{Proof:}

\leftskip=3ex

First of all, note that the operator ${\rm det}{\bf M}^{-}_{\cal
N}(H^{-}_{\cal N})$ annihilates $\phi^{-}_{n}$,
\begin{eqnarray}
{\rm det}{\bf M}^{-}_{\cal N}(H^{-}_{\cal N})\phi^{-}_{n}
=\sum_{m}\{{\rm det}{\bf M}^{-}_{\cal N}({\bf S}^{-})\}_{n,m}\phi^{-}_{m}=0.
\end{eqnarray}
This is because that ${\rm det}{\bf M}^{-}_{\cal N}({\bf S}^{-})$ is
identically zero by the Cayley-Hamilton theorem. 
From this, the form of ${\rm det}{\bf M}^{-}_{\cal N}(H^{-}_{\cal N})$ is
determined as 
\begin{eqnarray}
{\rm det}{\bf M}^{-}_{\cal N}(H^{-}_{\cal N})=F(p,q)P_{\cal N}, 
\label{sec_p_mh:f}
\end{eqnarray}
where $F(p,q)$ has the following form,
\begin{eqnarray}
F(p,q)=p^{\cal N}+f_{{\cal N}-1}(q)p^{{\cal N}-1}+\cdots+f_{1}(q)p+f_0(q). 
\label{sec_p_mh:f-form}
\end{eqnarray}
When we apply $H^{-}_{\cal N}$ to the above equation (\ref{sec_p_mh:f})
from the right, the right hand side becomes 
\begin{eqnarray}
F(p,q)P_{\cal N}H^{-}_{\cal N}= F(p,q)H^{+}_{\cal N}P_{\cal N},
\end{eqnarray}
and the left hand side becomes
\begin{eqnarray}
{\rm det}{\bf M}^{-}_{\cal N}(H^{-}_{\cal N})H^{-}_{\cal N}
=H^{-}_{\cal N}{\rm det}{\bf M}^{-}_{\cal N}(H^{-}_{\cal N})
=H^{-}_{\cal N}F(p,q)P_{\cal N}.
\end{eqnarray}
Thus we obtain 
\begin{eqnarray}
F(p,q)H^{+}_{\cal N}P_{\cal N}= H^{-}_{\cal N}F(p,q)P_{\cal N}.
\end{eqnarray}
Since any function can be written as $P_{\cal N}f(q)$, this equation means
\begin{eqnarray}
F(p,q)H^{+}_{\cal N}= H^{-}_{\cal N}F(p,q). 
\label{sec_p_mh:f-alg}
\end{eqnarray}
Thus if we define $p^{-\dagger}$ as
\begin{eqnarray}
p^{-\dagger}=P_{\cal N}^{\dagger}-F(p,q),  
\end{eqnarray}
it contains ${\cal N}-1$ derivatives with respect to $q$ at most and satisfies
\begin{eqnarray}
p^{-\dagger}H^{+}_{\cal N}= H^{-}_{\cal N}p^{-\dagger}.  
\end{eqnarray}
Using $p^{-}$, Eq.(\ref{sec_p_mh:f}) is rewritten as 
\begin{eqnarray}
{\rm det}{\bf M}^{-}_{\cal N}(H^{-}_{\cal N})=P^{\dagger}_{\cal N}P_{\cal N}
-p^{-\dagger}P_{\cal N}. 
\label{sec_p_mh:mh_oh1}
\end{eqnarray}
In a similar manner, the next equation can be shown, 
\begin{eqnarray}
{\rm det}{\bf M}^{+}_{\cal N}(H^{+}_{\cal N})=P_{\cal N}P^{\dagger}_{\cal N}
-p^{+} P^{\dagger}_{\cal N}, 
\label{sec_p_mh:mh_oh2}
\end{eqnarray}
where $p_K$ is an operator which contains ${\cal N}-1$ derivatives with
respect to $q$ at most
and satisfies 
\begin{eqnarray}
H^{+}_{\cal N}p^{+}= p^{+} H^{-}_{\cal N}.   
\end{eqnarray}
In terms of $P_{\cal N}$ and $P^{\dagger}_{\cal N}$, the Mother
Hamiltonian ${\cal H}_{\cal N}$ is given by 
\begin{eqnarray}
{\cal H}_{\cal N}=
\frac{1}{2}
\left(
\begin{array}{cc}
P_{\cal N}P^{\dagger}_{\cal N}&0\\
0& P^{\dagger}_{\cal N}P_{\cal N}
\end{array}
\right),  
\end{eqnarray}
thus if we define $q^{\pm}$ as
\begin{eqnarray}
q^{-}=p^{-\dagger}\psi
,\quad 
q^{+\dagger}=p^{+}\psi^{\dagger},
\end{eqnarray}
we obtain Eq.(\ref{sec_p_mh:mh_oh}) from Eqs.(\ref{sec_p_mh:mh_oh1})
and (\ref{sec_p_mh:mh_oh2}).

If the ${\cal N}$-fold supercharges are uniquely determined for given
$H_{\cal N}^{\pm}$,  $p^{\pm}$ must be zero
since $P_{\cal N}+p^{\pm}$ gives new ${\cal N}$-fold supercharges.   
Therefore, in this case we obtain Eq.(\ref{sec_p_mh:poly}).
 
\leftskip=0ex

\rightline{Q.E.D.}

It is worth noting that Eq.(\ref{sec_p_mh:poly})
can be more simplified as 
\begin{eqnarray}
{\cal H}_{\cal N}=\frac{1}{2}{\rm det}{\bf M}_{\cal N}^{+}({\bf H}_{\cal N}) 
=\frac{1}{2}{\rm det}{\bf M}_{\cal N}^{-}({\bf H}_{\cal N}). 
\end{eqnarray}
This will be proven in section \ref{sec_p_ps}.

\subsection{Spectrum}
\label{sec_gp_s}

Just as ordinary supersymmetry, 
bosonic states of ${\cal N}$-fold
supersymmetric systems and fermionic ones have one to one correspondence
unless the states are eigenstates of the Mother Hamiltonian with
zero eigenvalue.
To see this, let us consider a normalized bosonic state
$\Phi^{-}_n$ which satisfy 
\begin{eqnarray}
H^{-}_{\cal N}\Phi^{-}_n=E^{-}_n\Phi^{-}_n. 
\label{sec_p_s:hb}
\end{eqnarray}
Since ${\cal H}_{\cal N}$ commutes with ${\bf H}_{\cal N}$,  
$\Phi^{-}_n$ can
be simultaneously an eigenstate of the Mother Hamiltonian:
\begin{eqnarray}
{\cal H}_{\cal N}
\left(
\begin{array}{c}
0 \\
\Phi^{-}_{n}
\end{array}
\right)
={\cal E}_{n}
\left(
\begin{array}{c}
0 \\
\Phi^{-}_{n}
\end{array}
\right).
\label{sec_p_s:mhb}
\end{eqnarray}
If ${\cal E}_n$ is not zero, the following normalized state
$\Phi^{+}_n$ exists,
\begin{eqnarray}
\Phi^{+}_{n}\equiv \frac{P_{\cal N}}{\sqrt{{\cal E}_n}}\Phi^{-}_{n}. 
\label{sec_p_s:fs}
\end{eqnarray}
From Eq.(\ref{sec_dons:n-fold_alg2}), we can easily see that this state
is an eigenstate of the fermionic Hamiltonian $H^{+}_{\cal N}$ with the
same energy $E^{-}_n$; 
\begin{eqnarray}
H^{+}_{\cal N}\Phi^{+}_{n}=E^{-}_{n}\Phi^{+}_{n}.
\end{eqnarray}
Furthermore, this state is also the eigenstate of the Mother
Hamiltonian with the same ${\cal E}_n$;
\begin{eqnarray}
{\cal H}_{\cal N}
\left(
\begin{array}{c}
 \Phi^{+}_n\\
0
\end{array}
\right)
={\cal E}_n
\left(
\begin{array}{c}
\Phi^{+}_n \\
0
\end{array}
\right),
\end{eqnarray}
since ${\cal H}_{\cal N}$ commutes with $Q_{\cal N}^{\dagger}$ and 
\begin{eqnarray}
\left(
\begin{array}{c}
\Phi^{+}_n \\
0
\end{array} 
\right)
=\frac{Q_{\cal N}^{\dagger}}{\sqrt{{\cal E}_{\cal N}}}
\left(
\begin{array}{c}
0 \\
\Phi^{-}_n
\end{array}
\right).
\end{eqnarray}
In a similar manner, bosonic states can be
constructed from fermionic ones at each energy levels unless
${\cal E}_n=0$.

For states with ${\cal E}_n=0$, the eigenvalues of $H^{\pm}_{\cal
N}$ are determined algebraically.  
Bosonic states $\Phi^{-}_n$ with ${\cal E}_n=0$ satisfy
$P_{\cal N}\Phi^{-}_n=0$. 
Thus using Eq.(\ref{sec_p_mh:mh_oh}) (or more directly
Eq.(\ref{sec_p_mh:f})), 
we obtain 
\begin{eqnarray}
{\rm det}{\bf M}^{-}_{\cal N}(E^{-}_n)=0, 
\label{sec_e:men}
\end{eqnarray}
where $E^{-}_n$ is the eigenvalue of $H^{-}_{\cal N}$.
For fermionic states with ${\cal E}_n=0$, we obtain the following
algebraic equation in the same way,
\begin{eqnarray}
{\rm det}{\bf M}^{+}_{\cal N}(E^{+}_n)=0, 
\label{sec_e:nen}
\end{eqnarray}
where $E^{+}_n$ is the eigenvalue of $H^{+}_{\cal N}$.

\subsection{Polynomial supersymmetry}
\label{sec_p_ps}

A system is defined to have ${\cal N}$-th order polynomial
supersymmetry \cite{AIN} if the system is ${\cal N}$-fold supersymmetric
and its Mother Hamiltonian is given by an ${\cal N}$-th order polynomial
of Hamiltonian ${\bf H}_{\cal N}$. 
Here we show that if $H_{\cal N}^{-}\neq H_{\cal N}^{+}$,
any ${\cal N}$-fold supersymmetric system have ${\cal
M}$-th order polynomial supersymmetry with ${\cal M}\le {\cal N}$.

First consider the case that the ${\cal N}$-fold supersymmetric system
has a unique ${\cal N}$-fold supercharge for $H_{\cal N}^{\pm}$. 
In this case, the Mother Hamiltonian is given by Eq.(\ref{sec_p_mh:poly}).
Now we consider the following state
\begin{eqnarray}
\left(
\begin{array}{c}
 0\\
\Phi^{-}
\end{array}
\right), 
\end{eqnarray}
where $\Phi^{-}$ satisfies
\begin{eqnarray}
H^{-}_{\cal N}\Phi^{-}=E\Phi^{-}.
\end{eqnarray}
Here we do not require normalizability of $\Phi^{-}$ so $E$ may be an
arbitrary constant.
This state is also an eigenstate of the Mother Hamiltonian and the
eigenvalue ${\cal E}$ becomes 
\begin{eqnarray}
{\cal E}=\frac{1}{2}{\rm det}{\bf M}^{-}_{\cal N}(E).
\end{eqnarray}
Now we construct the following fermionic state
\begin{eqnarray}
\left(
\begin{array}{c}
P_{\cal N}\Phi^{-} \\
0
\end{array}
\right),  
\end{eqnarray}
and apply the Mother Hamiltonian to this.
Using
\begin{eqnarray}
H_{\cal N}^{+}P_{\cal N}\Phi^{-}= 
P_{\cal N}H_{\cal N}^{-}\Phi^{-}=EP_{\cal N}\Phi^{-}, 
\end{eqnarray}
we obtain
\begin{eqnarray}
\frac{1}{2}{\rm det}{\bf M}^{+}_{\cal N}(E)={\cal E}.
\end{eqnarray}
Eliminating ${\cal E}$ in the above equations, we obtain 
\begin{eqnarray}
{\rm det}{\bf M}^{-}_{\cal N}(E)={\rm det}{\bf M}^{+}_{\cal N}(E)
\end{eqnarray}
for any $E$.
Therefore, Eq.(\ref{sec_p_mh:poly}) becomes
\begin{eqnarray}
{\cal H}_{\cal N}
=\frac{1}{2} {\rm det}{\bf M}^{-}_{\cal N}({\bf H}_{\cal N})
=\frac{1}{2} {\rm det}{\bf M}^{+}_{\cal N}({\bf H}_{\cal N}).
\end{eqnarray}
Thus the system is ${\cal N}$-th order polynomial supersymmetric.

Next consider the case that the ${\cal N}$-fold supercharge is not
uniquely determined for given $H_{\cal N}^{\pm}$. 
As is shown in section \ref{sec_gp_mh}
the system has ${\cal N}_1$-fold supersymmetry with ${\cal N}_1<
{\cal N}$ in this case.
If this ${\cal N}_1$-fold supercharge is uniquely determined for given
$H_{\cal N}^{\pm}$, 
we can show in a manner similar to the above that the system has ${\cal
N}_1$-th polynomial supersymmetry. 
If this ${\cal N}_1$-fold supercharge is not uniquely determined, we
again obtain an ${\cal N}_2$-fold supercharge with ${\cal N}_2<{\cal
N}_1<{\cal N}$.  
If this ${\cal N}_2$-fold supercharge is uniquely determined, the system
has ${\cal N}_2$-th order polynomial supersymmetry.
We continue this procedure
until the obtained supercharge is uniquely determined or it becomes
$0$-th fold one.   
If the former is realized, the system is proved to have ${\cal N}_i$-th order
polynomial supersymmetry with ${\cal N}_i< {\cal N}$. 
If the latter is realized, there exist a function $\tilde{w}_0(q)$ which
satisfies
\begin{eqnarray}
\tilde{w}_0(q)H_{\cal N}^{-}=H_{\cal N}^{+}\tilde{w}_0(q) 
\label{sec_p_ps:0}
\end{eqnarray}
Comparing the first derivative terms in this equation, we find that 
$\tilde{w}_0$ does not depend on $q$.
Thus Eq.(\ref{sec_p_ps:0}) indicates that $H_{\cal N}^{-}=H_{\cal
N}^{+}$. 
This contradicts the assumption and
shows that the latter case is not realized.

\subsection{Generalized Witten index}
\label{sec_gp_gwi}

The Witten index of ordinary supersymmetry
can be generalized to ${\cal N}$-fold supersymmetric systems.
For polynomial supersymmetry, the generalization was first discussed in
Ref.\cite{AIS}. 
When the energy of the
Mother Hamiltonian is not zero (namely ${\cal E}_n\neq 0$), 
the bosonic and fermionic states form pairs.
Thus only states with ${\cal E}_n=0$ contribute to the Witten index
${\rm tr}(-1)^{F}$, 
\begin{eqnarray}
{\rm tr}(-1)^{F}={\rm dim}\,{\rm Kernel}\, Q_{\cal N}
-{\rm dim}\,{\rm Cokernel}\, Q_{\cal N}. 
\label{wittenindex}
\end{eqnarray}
The index takes integer values since the number of states with zero
energy of the Mother Hamiltonian is finite ($2{\cal N}$ at most). 
The expression (\ref{wittenindex}) shows that 
if this index is not zero, at least one ${\cal N}$-fold supersymmetric
state exists.

\subsection{Non-renormalization theorems}
\label{sec_gp_nt}

Non-renormalization theorems are characteristic features of
supersymmetric systems.  
The corresponding non-renormalization theorems also hold in ${\cal N}$-fold
supersymmetric systems. 
For example,  non-renormalization theorems hold for the
generalized Witten index.  
Because this index takes integer values,  it is also
an adiabatic invariant as well as the ordinary one and does not suffer
from quantum corrections.  
Furthermore, by an argument analogous to ordinary supersymmetry
\cite{Wit2}, we can show that perturbation theory does not break ${\cal
N}$-fold supersymmetry spontaneously.  

There exist other kinds of non-renormalization theorems in the
${\cal N}$-fold supersymmetric systems.
For states with ${\cal E}_n\neq 0$, 
the bosonic spectra and the fermionic ones are the same, thus the
perturbative corrections for them are also the same.
This property enables us to prove the non-renormalization of the energy
splittings for the ${\cal N}$-th and higher excited states of an
asymmetric double-well potential \cite{AKOSW2}.     
Furthermore, it was shown that in asymmetric double
well potentials \cite{AKOSW2} and periodic potentials \cite{ASTY}, the
perturbation series of the energies for states with ${\cal E}_n=0$
are convergent. 
This is because that ${\cal N}$-fold supersymmetry of these models
cannot be broken by any perturbative corrections.
The latter example of the non-renormalization theorem can be generalized to
a class of ${\cal N}$-fold supersymmetric systems, which will be
explained in section \ref{sec_e_typea}.

\subsection{Quasi-solvability and ${\cal N}$-fold supersymmetry}
\label{sec_gp_qans}

In closing this section, we note a close relationship between
quasi-solvability and ${\cal N}$-fold supersymmetry.
For a finite order differential operator $P$, let us
consider a function $\phi$ which satisfies $P\phi=0$.
A system with a Hamiltonian $H$ is defined to be ``quasi-solvable'' if
$PH\phi=0$ holds for any such $\phi$s.
Namely, if a system is quasi-solvable, the space ${\cal V}$ defined by 
${\cal V}=\{\phi| P\phi=0\}$ is closed by the action of $H$. 
If we introduce the basis $\phi_n$ of ${\cal V}$, we obtain 
\begin{eqnarray}
H\phi_n=\sum_{n=1}^{{\rm dim}{\cal V}} {\bf S}_{n,m}\phi_m 
\label{sec_p_ps:sv}
\end{eqnarray} 
This means that a part of the spectra of the quasi-solvable system can
be solved by the characteristic equations for ${\bf S}$ 
which is finite dimensional.

For example, ${\cal N}$-fold supersymmetric systems are quasi-solvable.
The projective operators for $H^{-}_{\cal N}$ and $H^{+}_{\cal N}$ are
$P_{\cal N}$ and $P_{\cal N}^{\dagger}$ respectively. 
A part of the spectra of the systems is solved by the following  
algebraic equations,
\begin{eqnarray}
{\rm det}{\bf M}^{-}_{\cal N}(E^{-}_n)=0,
\quad
{\rm det}{\bf M}^{+}_{\cal N}(E^{+}_n)=0,
\label{sec_p_qs:ae}
\end{eqnarray}
where $E_n^{-}$ and $E_n^{+}$ are eigenvalues of $H_{\cal N}^{-}$ and
$H_{\cal N}^{+}$ respectively.

This quasi-solvability for ${\cal N}$-fold supersymmetric systems
comes from the ${\cal N}$-fold supersymmetric algebra, but the converse
is also true: 
If a system is quasi-solvable and $P$ is an ${\cal N}$-th order
differential operator, it also becomes ${\cal N}$-fold
supersymmetric.

\noindent
\underline{Proof:}

\leftskip=3ex
We assume that the projective operator $P$ and the Hamiltonian $H$ have
the following form, 
\begin{eqnarray}
P&=&p^{\cal N}+c_{{\cal N}-1}(q)p^{{\cal N}-1}+\cdots
+c_1(q)p+c_0(q),
\nonumber\\
H&=&\frac{1}{2}p^2+V(q).
\end{eqnarray}
For this $P$ and $H$, we introduce another Hamiltonian $K$ as follows,
\begin{eqnarray}
K=\frac{1}{2}p^2+U(q)
,\quad 
U(q)=V(q)+ic_{{\cal N}-1}'(q).
\end{eqnarray}
If we introduce the operator 
$G(p,q)\equiv PH-KP$,
it contains ${\cal N}-1$ derivatives with respect to $q$ at most and
$G(p,q)\phi=0$ for any $\phi$ which satisfy $P\phi=0$. 
But as operators which contain ${\cal N}-1$ derivatives at most cannot
annihilate ${\cal N}$ independent functions 
non-trivially, this means that $G(p,q)\equiv 0$.
Therefore, if we identify $P_{\cal N}=P$, $H^{-}_{\cal N}=H$ and $H^{+}_{\cal
N}=K$, we obtain an ${\cal N}$-fold supersymmetric system.

\leftskip=0ex
\rightline{Q.E.D.}

Note that all the eigenvalues of ${\bf S}$ in Eq.(\ref{sec_p_ps:sv}) are not
necessarily physical ones.
This is because the quasi-solvability does not require that ${\cal V}$ is
a quantum physical space, that is, ${\rm L}^2$.
When ${\cal V}$ is ${\rm L}^2$, 
the system is often called  ``quasi-exactly solvable'' \cite{Tur,Shi,Ush}.
In this case, all the eigenvalues are physical.
Even when the elements of ${\cal V}$ are not normalizable, 
if they become normalizable in the
all order of the perturbation theory, the eigenvalues of ${\bf S}$ are
exact in the perturbation theory. 
We dub this ``quasi-perturbatively solvable''.

Among the known ${\cal N}$-fold symmetric models,
the quartic model found in \cite{AKOSW2} is 
quasi-perturbatively solvable, while 
the periodic one in \cite{ASTY} and
the sextic one in \cite{AST} are quasi-exactly solvable,
the exponential one in \cite{AST,KP1} can be either of those,
depending on a parameter.
In the perturbation theory, all the models have 
normalizable eigenstates of Eq.(\ref{sec_p_qs:ae}) 
which are ${\cal N}$-fold supersymmetric. 
In the quasi-exactly solvable models, they remain normalizable 
even if non-perturbative effects are taken into account.
Thus the physical states in this type of models 
contain ${\cal N}$-fold supersymmetric ones.
But, in the quartic model, 
these states are no longer normalizable
if non-perturbative effects are taken into account.
Thus the physical states in the latter model 
do not contain ${\cal N}$-fold supersymmetric ones.

Special cases of the correspondence between quasi-solvability 
and ${\cal N}$-fold supersymmetry were previously reported;
for the quartic potential in Ref.\cite{AKOSW2}, 
for the periodic potential in Ref.\cite{ASTY},
for the exponential potentials in Ref.\cite{KP1}, 
for the sextic potential in Ref.\cite{DDT}.
In this subsection, we have proved that the correspondence is general
and does not rely on any specific models.

\section{Examples}
\label{e}

\subsection{2-fold supersymmetry}
\label{sec_e_2s}

The first example of ${\cal N}$-fold supersymmetric models is the 2-fold
supersymmetric one.
Under the assumption that the Mother Hamiltonian ${\cal H}_2$ becomes a
polynomial of ${\bf H}_2$, the 2-fold supersymmetric model was first
constructed in Ref.\cite{AICD}.
Here we do not assume this.

In general, the 2-fold supercharges are given by 
\begin{eqnarray}
P_2=p^2+w_1(q)p+w_0(q). 
\end{eqnarray}
To be 2-fold supersymmetric, the following relation must hold,
\begin{eqnarray}
P_2H^{-}_2-H^{+}_2P_2=0. 
\end{eqnarray}
Since the left hand side of the above is given by
\begin{eqnarray}
&&2(P_2H^{-}_2-H^{+}_2P_2)
\nonumber\\
&&=(-\partial^2-iw_1\partial+w_0)(-\partial^2+2V_2^{-})
-(-\partial^2+2V_2^{+})(-\partial^2-iw_1\partial+w_0)
\nonumber\\
&&=-2(V_2^{-}+iw_1'-V_2^{+})\partial^2
+(-4V_2^{-}{}'-2iw_1V_2^{-}-iw_1''+2w_0'+2iw_1V_2^{+})\partial
\nonumber\\
&&\quad-2V_2^{-}{}''-2iw_1V_2^{-}{}'+2w_0V_2^{-}+w_0''-2w_0V_2^{+}, 
\end{eqnarray}
the following three equations have to be satisfied,
\begin{eqnarray}
&&V_2^{+}-V_2^{-}=iw_1', 
\label{sec_e:2fold1}
\\
&&iw_1''-2iw_1(V_2^{+}-V_2^{-})-2w_0'+4V_2^{-}{}'=0,
\label{sec_e:2fold2}
\\
&&w_0''-2w_0(V_2^{+}-V_2^{-})-2iw_1V_2^{-}{}'-2V_2^{-}{}''=0.
\label{sec_e:2fold3}
\end{eqnarray}
Eliminating $V_2^{-}$ and $V_2^{+}$ from these equations, we
obtain
\begin{eqnarray}
-iw_1w_0'-2iw_1'w_0
+\frac{1}{2}\left(iw_1'''+w_1''w_1+2w_1'{}^2+2iw_1'w_1^2\right)=0. 
\label{sec_e:w0w1}
\end{eqnarray}
This equation is easily solved if we introduce the following function
$\Omega(q)$,  
\begin{eqnarray}
\Omega=-iw_0w_1^2. 
\end{eqnarray}
From Eq.(\ref{sec_e:w0w1}), the function $\Omega(q)$ satisfies
\begin{eqnarray}
\Omega'=-\frac{1}{2}
\left(iw_1'''w_1+w_1''w_1^2+2w_1'{}^2w_1+2iw_1'w_1^3\right),  
\end{eqnarray}
thus 
\begin{eqnarray}
\Omega=-\frac{1}{2}\left(iw_1''w_1-\frac{1}{2}iw_1'{}^2+w_1'w_1{}^2
+\frac{1}{2}iw_1^4+C\right), 
\end{eqnarray}
where $C$ is an arbitrary constant. 
So if $w_1(q)\not{\hspace{-0.6ex}\equiv}\,\, 0$, $w_0(q)$ is given as follows, 
\begin{eqnarray}
w_0(q)=\frac{1}{4}w_1(q)^2+\frac{1}{2}
\left(\frac{w_1''(q)}{w_1(q)}
-\frac{w_1'(q)^2}{2w_1(q)^2}
-\frac{iC}{w_1(q)^2}\right)
-\frac{1}{2}iw_1'(q). 
\end{eqnarray}
The potentials $V_2^{\pm}$ can be also written in terms of $w_1$.
Eliminating $V_2^{+}$ from Eqs.(\ref{sec_e:2fold1}) and
(\ref{sec_e:2fold2}), we obtain  
\begin{eqnarray}
\left(iw_1'+w_1^2-2w_0+4V_2^{-}\right)'=0. 
\end{eqnarray}
So if we omit irrelevant integral constants, $V_2^{-}(q)$ is
given by
\begin{eqnarray}
V_2^{-}&&= -\frac{1}{4}\left(iw_1'+w_1^2-2w_0\right)
\nonumber\\
&&=-\frac{1}{8}w_1(q)^2+\frac{1}{4}
\left(\frac{w_1''(q)}{w_1(q)}
-\frac{w_1'(q)^2}{2w_1(q)^2}
-\frac{iC}{w_1(q)^2}\right) 
-\frac{1}{2}iw_1'(q).
\end{eqnarray}
The remaining potential $V_2^{+}$ is obtained by Eq.(\ref{sec_e:2fold1}). 

When $w_1(q)\equiv 0$, the above solution is not valid.
In this case, however, Eqs.(\ref{sec_e:2fold1})--(\ref{sec_e:2fold3})
reduce to the following simple equations, 
\begin{eqnarray}
V_2^{+}-V_2^{-}=0, 
\quad 
-2w_0'+4V_2^{-}{}'=0,
\quad
w_0''-2V_2^{-}{}''=0,  
\end{eqnarray}
thus the solution is easily obtained as
\begin{eqnarray}
V_2^{+}=V_2^{-}=\frac{1}{2}w_0.
\end{eqnarray}
This solution is trivial and useless,
since the supercharge $P_2$ coincides with the Hamiltonians
$H^{\pm}_2$.

In summary, the non-trivial 2-fold supersymmetric system is generally
given as follows:
\begin{eqnarray}
&&P_2=-\partial^2-iw_1(q)\partial+w_0(q), 
\\
&&w_0(q)=\frac{1}{4}w_1(q)^2+\frac{1}{2}
\left(\frac{w_1''(q)}{w_1(q)}
-\frac{w_1'(q)^2}{2w_1(q)^2}
-\frac{iC}{w_1(q)^2}\right)
-\frac{1}{2}iw_1'(q), 
\\
&&V_2^{\pm}
=-\frac{1}{8}w_1(q)^2+\frac{1}{4}
\left(\frac{w_1''(q)}{w_1(q)}
-\frac{w_1'(q)^2}{2w_1(q)^2}
-\frac{iC}{w_1(q)^2}\right) 
\pm\frac{1}{2}iw_1'(q).
\end{eqnarray}

For given $V_2^{-}$ and $V_2^{+}$, the above 2-fold supercharges
are determined uniquely unless 
\begin{eqnarray}
w_1''-2iw_1w_1'-2iV_2^{-}{}'\propto w_1' 
\end{eqnarray}
holds.
To see this, we introduce another 2-fold supercharges which are given by 
substitution of the following $\hat{P}_2$ for $P_2$: 
\begin{eqnarray}
\hat{P}_2=-\partial^2-i\hat{w}_1(q)\partial+\hat{w}_0(q). 
\end{eqnarray}
$\hat{P}_2$ also satisfies $\hat{P}_2 H^{-}_2-H^{+}_2\hat{P}_2=0$.
If we define $\Delta w_i=w_i-\hat{w}_i$ $(i=0,1)$, they satisfy
\begin{eqnarray}
&&i\Delta w_1'=0,
\label{sec_e:p2unique1}
\\
&&i\Delta w_1''-2i\Delta w_1(V_2^{+}-V_2^{-})-2\Delta w_0'=0,
\label{sec_e:p2unique2}
\\
&&\Delta w_0''-2\Delta w_0(V_2^{+}-V_2^{-})-2i\Delta w_1 V_2^{-}{}'=0.
\label{sec_e:p2unique3}
\end{eqnarray}
From the first equation (\ref{sec_e:p2unique1}), $\Delta w_1$ is
determined as
\begin{eqnarray}
\Delta w_1=C_1, 
\end{eqnarray}
where $C_1$ is a constant. Substituting this for Eq.(\ref{sec_e:p2unique2}),
we obtain
\begin{eqnarray}
&&2C_1w_1'-2\Delta w_0'=0, 
\end{eqnarray}
so $\Delta w_0$ becomes
\begin{eqnarray}
\Delta w_0= C_1 w_1+C_2,
\end{eqnarray}
where $C_2$ is a constant.
Thus Eq.(\ref{sec_e:p2unique3}) becomes
\begin{eqnarray}
C_1 (w_1''-2iw_1w_1'-2iV_2^{-}{}')=2iC_2 w_1'.
\end{eqnarray}
Unless
\begin{eqnarray}
w_1''-2iw_1w_1'-2iV_2^{-}{}'\propto w_1', 
\end{eqnarray}
only solution of this equation is $C_1=C_2=0$, and this means
that $\hat{w}_0=w_0$ and $\hat{w}_1=w_1$.

\subsection{Type A ${\cal N}$-fold supersymmetry}
\label{sec_e_typea}

For the second example of ${\cal N}$-fold supersymmetry, we consider
a particular class of ${\cal N}$-fold supercharges which we call {\em
type A} \cite{AST}.
The form of the type A ${\cal N}$-fold supercharges $P_{\cal N}$
is defined as follows: 
\begin{eqnarray}
P_{\cal N}
&&=
\left(D+i({\cal N}-1)E(q)\right)
\left(D+i({\cal N}-2)E(q)\right)
\cdots
\left(D+iE(q)\right)
D
\nonumber\\
&&\equiv\prod_{k=0}^{{\cal N}-1}(D+ikE(q)),  
\end{eqnarray}
where $D=p-iW(q)$.
The ${\cal N}$-fold supersymmetric models considered in
Refs.\cite{AKOSW2,ASTY} are in this class.
A type A model was also considered in Refs.\cite{DDT,KP2}.  

For this class of ${\cal N}$-fold supercharges, a system is ${\cal
N}$-fold supersymmetric when the following conditions are satisfied: 
\begin{eqnarray}
V^{\pm}_{\cal N}
&&=\frac{1}{2}(W^2+v_{\cal N}^{\pm}),
\nonumber\\
v^{\pm}_{\cal N}
&&=-({\cal N}-1)E(q)W(q)
+\frac{({\cal N}-1)(2{\cal N}-1)}{6}E(q)^2
\nonumber\\
&&\quad -\frac{{\cal N}^2-1}{6}E'(q)
\pm{\cal N}\left(W'(q)-\frac{{\cal N}-1}{2}E'(q)\right).
\label{sec_e:typeaa}
\\
W(q)&&=\frac{1}{2}E(q)+Ce^{-\int^{q}dq_1E(q_1)}\int^{q}dq_2
\left(e^{\int^{q_2}dq_3 E(q_3)}\int^{q_2} dq_4 e^{\int^{q_4} dq_5 E(q_5)}
\right)
\quad ({\cal N}\ge 2),
\nonumber\\
\label{sec_e:typeab}
\\
E'''(q)&&+E(q)E''(q)+2E'(q)^2-2E(q)^2E'(q)=0
\quad ({\cal N}\ge 3),
\label{sec_e:typeac}
\end{eqnarray}
where $C$ is an arbitrary constant.

\noindent
\underline{Proof:}

\leftskip=3ex

We prove the above conditions (\ref{sec_e:typeaa})--(\ref{sec_e:typeac})
inductively. 
For ${\cal N}=1$, Eqs.(\ref{sec_e:typeaa})--(\ref{sec_e:typeac}) reduce to
\begin{eqnarray}
V_1^{\pm}=\frac{1}{2}(W^2\pm W'), 
\end{eqnarray}
which is the ordinary supersymmetric case.
Thus the system is ${\cal N}$-fold supersymmetric in this case.  
Next, we suppose that the conditions
(\ref{sec_e:typeaa})--(\ref{sec_e:typeac}) hold for an integer ${\cal N}$.
This assumption implies that the ${\cal N}$-fold superalgebra 
$P_{\cal N}H^{-}_{\cal N}=H^{+}_{\cal N}P_{\cal N}$ holds in this case.
Then, if we put 
\begin{eqnarray}
H^{+}_{{\cal N}+1}=H^{+}_{\cal N}+h^{+}_{\cal N},
\quad 
H^{-}_{{\cal N}+1}=H^{-}_{\cal N}+h^{-}_{\cal N},
\end{eqnarray}
and use the relation $P_{\cal N}H^{-}_{\cal N}=H^{+}_{\cal N}P_{\cal N}$, we obtain 
\begin{eqnarray}
P_{{\cal N}+1}H^{-}_{{\cal N}+1}-H^{+}_{{\cal N}+1}P_{{\cal N}+1} 
=[D+i{\cal N}E, H^{+}_{\cal N}]P_{\cal N}
-h^{+}_{\cal N}P_{{\cal N}+1}+P_{{\cal N}+1}h^{-}_{\cal N}.
\label{sec_e:ph-kp}
\end{eqnarray}
To facilitate the following calculation, we introduce $U$ as follows
\begin{eqnarray}
U(q)=e^{\int^{q}dq'W(q')}.
\end{eqnarray}
Then the Hamiltonian $H^{+}_{\cal N}$ and the supercharge $P_{\cal N}$
are rewritten as
\begin{eqnarray}
UH^{+}_{\cal N}U^{-1}&&=\frac{1}{2}(-\partial^2+2W\partial+W'+v_{\cal N}^{+}), 
\nonumber\\
UP_{\cal N}U^{-1}
&&=(-i)^{\cal N}\left(\partial-({\cal N}-1)E(q)\right) 
\left(\partial-({\cal N}-2)E(q)\right)\cdots
\left(\partial-E(q)\right)\partial
\nonumber\\
&&\equiv (-i)^{\cal N}\prod_{k=0}^{{\cal N}-1}\left(\partial -k E(q)\right)
\equiv (-i)^{\cal N}\tilde{P}_{\cal N}.  
\end{eqnarray}
Now Eq.(\ref{sec_e:ph-kp}) is calculated as
\begin{eqnarray}
I_{{\cal N}+1}
&&\equiv 2 i^{{\cal N}+1}U
\left(P_{{\cal N}+1}H^{-}_{{\cal N}+1}-H^{+}_{{\cal N}+1}P_{{\cal N}+1}
\right) U^{-1}
\nonumber\\
&&=[\partial -{\cal N}E,-\partial^2+2W\partial+W'+v^{+}_{\cal N}]
\tilde{P}_{\cal N}
-2h^{+}_{\cal N}\tilde{P}_{{\cal N}+1}
+2\tilde{P}_{{\cal N}+1}h^{-}_{\cal N}
\nonumber\\
&&=2\left(W'-{\cal N}E'-h^{+}_{\cal N}+h^{-}_{\cal N}\right)
\partial\tilde{P}_{\cal N}
\nonumber\\
&&\quad +\left(
v^{+}_{\cal N}{}'+W''-{\cal N}E''+2{\cal N}E'W+2{\cal N}
(h^{+}_{\cal N}-h^{-}_{\cal N})E
\right)\tilde{P}_{\cal N}
+2[\tilde{P}_{{\cal N}+1}, h^{-}_{\cal N}].
\nonumber\\
\label{sec_e:in+1}
\end{eqnarray}
From Eq.(\ref{sec_e:in+1}), we see that $I_{{\cal N}+1}$ contains up to
$({\cal N}+1)$-th derivatives.
Therefore, $I_{{\cal N}+1}=0$ if and only if all the coefficients of
$\partial^k \,(k=0,1,\cdots,{\cal N}+1)$ vanish.
The $\partial^{{\cal N}+1}$ term comes only from the first term of the
right hand side of Eq.(\ref{sec_e:in+1}) and thus
\begin{eqnarray}
h^{+}_{\cal N}-h^{-}_{\cal N}=W'-{\cal N}E'. 
\label{sec_e:typea1}
\end{eqnarray}
When this condition (\ref{sec_e:typea1}) is satisfied, $I_{{\cal N}+1}$
now reads 
\begin{eqnarray}
I_{{\cal N}+1}
&&=\left(v^{+}_{\cal N}{}'+W''-{\cal N}E''+2{\cal N}E'W+2{\cal N}EW'
-2{\cal N}^2EE'\right)\tilde{P}_{\cal N} 
+2[\tilde{P}_{{\cal N}+1},h_{\cal N}]
\nonumber\\
&&=\left(v^{+}_{\cal N}{}'+W''-{\cal N}E''+2{\cal N}E'W+2{\cal N}EW'
-2{\cal N}^2EE'\right)\tilde{P}_{\cal N}
\nonumber\\ 
&&\quad +2h^{-}_{\cal N}{}'\tilde{P}_{\cal N}
+2\left(\partial -{\cal N}E\right)
[\tilde{P}_{\cal N},h^{-}_{\cal N}]
\nonumber\\
&&=\left(v^{+}_{\cal N}{}'+W''-{\cal N}E''+2{\cal N}E'W+2{\cal N}EW'
-2{\cal N}^2EE'\right)\partial^{\cal N}
\nonumber\\
&& \quad +2h^{-}_{\cal N}{}'\partial^{\cal N}
+2{\cal N}h^{-}_{\cal N}{}'\partial^{\cal N}+O(\partial ^{{\cal N}-1}).
\label{sec_e:in+12}
\end{eqnarray}
To eliminate the $\partial^{\cal N}$ term, the following
condition have to be satisfied, 
\begin{eqnarray}
2({\cal N}+1)h^{-}_{\cal N}{}'=
-\left(v^{+}_{\cal N}{}'+W''-{\cal N}E''+2{\cal N}E'W+2{\cal N}EW'
-2{\cal N}^2EE'\right). 
\end{eqnarray}
From this equation, we obtain
\begin{eqnarray}
 h^{-}_{\cal N}
&&=-\frac{1}{2({\cal N}+1)}\left(v^{+}_{\cal N}+W'-{\cal N}E'
+2{\cal N}EW-{\cal N}^2E^2\right)
\nonumber\\
&&=\frac{1}{2}\left[-EW+\frac{4{\cal N}-1}{6}E^2
-\frac{2{\cal N}+1}{6}E'-(W'-{\cal N}E')\right]
\label{sec_e:typea2}
\end{eqnarray}
Here we omit an irrelevant constant which only affects the origin of
the energy.
Combining Eqs.(\ref{sec_e:typeaa}), (\ref{sec_e:typea1}) and this, we
finally find
\begin{eqnarray}
v^{\pm}_{{\cal N}+1}
&&=v^{\pm}_{\cal N}+2h^{\pm}_{\cal N}
\nonumber\\
&&=-{\cal N}EW+\frac{{\cal N}(2{\cal N}+1)}{6}E^2
-\frac{{\cal N}({\cal N}+2)}{6}E'\pm({\cal N}+1)
\left(W'-\frac{{\cal N}}{2}E'\right), 
\end{eqnarray}
which are nothing but the assumed form of $v_{\cal N}^{+}$ and
$v_{\cal N}^{-}$ with ${\cal N}$ replaced with ${{\cal N}+1}$. 
When we use the condition (\ref{sec_e:typea2}), from the second line of
the right hand side of Eq.(\ref{sec_e:in+12}), $I_{{\cal N}+1}$ becomes
\begin{eqnarray}
I_{{\cal N}+1}
&&=-2{\cal N}h^{-}_{\cal N}{}'\tilde{P}_{\cal N}
+2\left(\partial -{\cal N}E\right) [\tilde{P}_{\cal N},h^{-}_{\cal N}]
\nonumber\\
&&={\cal N}({\cal N}+1)\left(h^{-}_{\cal N}{}''-Eh^{-}_{\cal N}{}'\right)
\partial^{{\cal N}-1}+O(\partial^{{\cal N}-2}).
\label{sec_e:in+13}
\end{eqnarray}
Thus we obtain
\begin{eqnarray}
h^{-}_{\cal N}{}''-Eh^{-}_{\cal N}{}'=0 .
\label{sec_e:typea3}
\end{eqnarray}
From Eq.(\ref{sec_e:typea2}), this equation becomes
\begin{eqnarray}
&&\left[\left(W-\frac{1}{2}E\right)'+E\left(W-\frac{1}{2}E\right)\right]''
-E\left[\left(W-\frac{1}{2}E\right)'+E\left(W-\frac{1}{2}E\right)\right]' 
\nonumber\\
&&\quad -\frac{2({\cal N}-1)}{3}\left[(E'+E^2)''-E(E'+E^2)'\right]=0.
\label{sec_e:typea32}
\end{eqnarray} 
This equation leads to Eqs.(\ref{sec_e:typeab}) and (\ref{sec_e:typeac}).
Once Eq.(\ref{sec_e:typea3}) holds, we can prove $I_{{\cal N}+1}=0$, by
using the following relation,
\begin{eqnarray}
[\tilde{P}_{\cal N}, h_{\cal N}^{-}]
={\cal M}h_{\cal N}^{-}{}'\tilde{P}_{{\cal N}-1} 
+\left[\prod_{k={\cal M}}^{{\cal N}-1}(\partial-kE),h_{\cal N}^{-}\right]
\tilde{P}_{\cal M} 
\quad
(0\le{\cal M}\le{\cal N}),
\label{sec_e:ph}
\end{eqnarray}
where $\tilde{P}_0$ and $\prod_{k=\cal N}^{{\cal N}-1}(\partial -kE)$ should be
understood as $\tilde{P}_0=1$ and $\prod_{k=\cal N}^{{\cal
N}-1}(\partial -kE)=0$. 
This relation is easily obtained by the next relation,
\begin{eqnarray}
\left(\partial-kE\right)h_{\cal N}^{-}{}'
=h_{\cal N}^{-}{}'\left(\partial-(k-1)E\right). 
\end{eqnarray}
Applying the relation (\ref{sec_e:ph}) with ${\cal M}={\cal N}$ to the
first line of the right hand side of Eq.(\ref{sec_e:in+13}), we
immediately find that $I_{{\cal N}+1}=0$.

\leftskip=0ex
\rightline{Q.E.D.}

For illustration, we give here two examples of the solutions of
Eqs. (\ref{sec_e:typeab}) and (\ref{sec_e:typeac}).  
The first one is 
\begin{eqnarray}
&&W(q)=\sin(q)+\frac{{\cal N}-1}{2}i,\\
&&E(q)=i.
\end{eqnarray}
The potentials of this system are \cite{ASTY}
\begin{eqnarray}
V_{\cal N}^{\pm}=\frac{1}{2}\sin^2(q)\pm \frac{\cal N}{2}\cos(q),
\end{eqnarray}
where we have omitted an irrelevant constant. 
This system is periodic and may be defined on a finite
region, $q\in ( 0,2\pi]$.
The second one is
\begin{eqnarray}
&&W(q)=C_1 q^3+C_2 q +\frac{2C_3-{\cal N}+1}{2q},\\
&&E(q)=\frac{1}{q},\\ 
&&V_{\cal N}^{\pm}=\frac{1}{2}w(q)^2
+\frac{(2C_3\mp{\cal N}-1)(2C_3\mp{\cal N}+1)}{8q^2}
\pm\left(\frac{\cal N}{2}\pm\frac{C_3}{3}\right)w'(q)
+\frac{2}{3}C_2C_3,
\end{eqnarray}
where $C_i$ $(i=1,2,3)$ are arbitrary constants and $w(q)=C_1 q^3+C_2 q$.
This is a sextic anharmonic oscillator with centrifugal
like potential and thus may be naturally defined on the positive
axis with proper boundary conditions at $q=0$ and $\infty$.

For type A ${\cal N}$-fold supersymmetric models, the matrices ${\bf
S}^{\pm}$ defined in section \ref{sec_gp_mh} can
be given explicitly. 
To see this, we define the following functions $\phi^{-}_n(q)$,
\begin{eqnarray}
\phi^{-}_n=(h)^{n-1}U^{-1}
\quad(n=1,\cdots,{\cal N}),
\end{eqnarray}
where $h$ is a function which satisfy
\begin{eqnarray}
h''-Eh'=0. 
\label{sec_e:h}
\end{eqnarray}
Integrating Eq.(\ref{sec_e:h}), we obtain 
\begin{eqnarray}
h(q)=c_1\int_{0}^{q}dq_1e^{\int^{q_1}_{0}dq_2E(q_2)}+c_2,
\label{sec_e:hc}
\end{eqnarray}
where $c_1$ and $c_2$ are arbitrary constants.
Because of Eq.(\ref{sec_e:h}), the functions $\phi^{-}_n$ satisfy
\begin{eqnarray}
P_{\cal M}\phi^{-}_n=hP_{\cal M}\phi^{-}_{n-1}+{\cal M}(-i)h'P_{{\cal M}-1}\phi^{-}_{n-1}, 
\end{eqnarray}
where $\phi^{-}_0$ and $P_{0}$ should be understood as $\phi^{-}_0=0$ and
$P_{0}=1$. 
From this, we obtain
\begin{eqnarray}
P_{\cal M}\phi^{-}_n=0  
\quad ({\cal M}\ge n ),
\label{sec_e:pp0}
\end{eqnarray}
thus all the $\phi^{-}_n$s satisfy 
$P_{\cal N}\phi^{-}_n=0$.
And if $h'\not{\hspace{-0.6ex}\equiv}\,\,0$, the functions $\phi^{-}_n$
are linearly independent from each other since the next equation holds, 
\begin{eqnarray}
P_{\cal M}\phi^{-}_{{\cal M}+1}={\cal M}!(-i)^{\cal M}h'{}^{\cal M}U^{-1}. 
\label{sec_e:ppneq0}
\end{eqnarray}
When $h'\equiv 0$, the independence is broken, thus 
we choose $h$ as it satisfies $h'\not{\hspace{-0.6ex}\equiv}\,\, 0$ in the
following. 

Using $\phi^{-}_n$, we define ${\bf S}^{-}$ by
\begin{eqnarray}
H^{-}_{\cal N}\phi^{-}_n=\sum_m {\bf S}^{-}_{n,m}\phi^{-}_m. 
\label{sec_e:sa}
\end{eqnarray}
Applying $P_{{\cal N}-1}$ to the both sides of the above equation
and using Eq.(\ref{sec_e:pp0}), we
obtain the following equation,
\begin{eqnarray}
P_{{\cal N}-1}H^{-}_{\cal N}\phi^{-}_n={\bf S}^{-}_{n,{\cal N}}
P_{{\cal N}-1}\phi^{-}_{\cal N},  
\end{eqnarray}
thus ${\bf S}^{-}_{n,{\cal N}}$ is determined as
\begin{eqnarray}
{\bf S}^{-}_{n,{\cal N}}=\frac{P_{{\cal N}-1}H^{-}_{\cal N}\phi^{-}_{n}}
{P_{{\cal N}-1}\phi^{-}_{\cal N}}. 
\label{sec_e:sncn}
\end{eqnarray}
The other elements of ${\bf S}^{-}$ are
determined inductively. 
They are given as follows,
\begin{eqnarray}
{\bf S}^{-}_{n,{\cal N}-m}=\frac{P_{{\cal N}-m-1}
\left(H^{-}_{\cal N}\phi^{-}_n-\sum_{k={\cal N}-m+1}^{{\cal N}}
{\bf S}^{-}_{n,k}\phi^{-}_k\right) }
{P_{{\cal N}-m-1}\phi^{-}_{{\cal N}-m}},
\label{sec_e:sncn-m}
\end{eqnarray}
where $n=1,\cdots,{\cal N}$ and $m=1,\cdots,{\cal N}-1$.

The matrix ${\bf S}^{+}_{n,m}$ are given in a similar manner as ${\bf
S}^{-}_{n,m}$. 
First note that $P_{\cal N}$ and its
hermitian conjugate $P_{\cal N}^{\dagger}$ are related by
\begin{eqnarray}
P_{\cal N}^{\dagger}=U^2VP_{\cal N}V^{-1}U^{-2}, 
\end{eqnarray}
where $V$ is defined by
\begin{eqnarray}
V(q)=e^{-\int^{q} dq'({\cal N}-1)E(q')}. 
\end{eqnarray}
Thus instead of $\phi_n^{-}$, we introduce the following functions
$\phi^{+}_n$, 
\begin{eqnarray}
\phi^{+}_n=U^2V\phi^{-}_n=(h)^{n-1}VU, \quad (n=1,\cdots,{\cal N}),  
\end{eqnarray}
and define the matrices ${\bf S}^{+}_{n,m}$ as follows,
\begin{eqnarray}
H^{+}_{\cal N}\phi^{+}_n=\sum_m {\bf S}^{+}_{n,m}\phi^{+}_m. 
\end{eqnarray}
${\bf S}^{+}_{n,m}$ is determined inductively as follows, 
\begin{eqnarray}
&&{\bf S}^{+}_{n,{\cal N}}
=\frac{P^{\dagger}_{{\cal N}-1}H^{+}_{\cal N}\phi^{+}_{n}}
{P^{\dagger}_{{\cal N}-1}\phi^{+}_{\cal N}}, 
\\
&&{\bf S}^{+}_{n,{\cal N}-m}=\frac{P^{\dagger}_{{\cal N}-m-1}
\left(H^{+}_{\cal N}\phi^{+}_n-\sum_{k={\cal N}-m+1}^{{\cal N}}
{\bf S}^{+}_{n,k}\phi^{+}_k\right) }
{P^{\dagger}_{{\cal N}-m-1}\phi^{+}_{{\cal N}-m}},
\end{eqnarray}
where $n=1,\cdots,{\cal N}$ and $m=1,\cdots, {\cal N}-1$.

A kind of non-renormalization theorem found
in Refs.\cite{AKOSW2,ASTY} can be generalized to all the type A models
which have $q_0$ such as $W(q_0)=0$. 
By redefinition of the origin of the coordinate $q$, we first set $q_0=0$. 
Then we introduce a coupling constant $g$ as
follows, 
\begin{eqnarray}
W(q)=\frac{1}{g}w(gq),
\quad
E(q)=ge(gq).
\label{sec_e:defg}
\end{eqnarray}
In the leading order of $g$, the potentials $V_{\cal N}^{\pm}$ 
become harmonic ones with
frequency $|w'(0)|$, 
\begin{eqnarray}
V_{\cal N}^{\pm}=\frac{1}{2}w'(0){}^2q^2+O(g).
\label{sec_e:vlg}
\end{eqnarray}
The following non-renormalization theorem holds for the first ${\cal N}$
excited states of either of these harmonic potentials $V_{\cal N}^{\pm}$:
If $w'(0)>0$, perturbative corrections for the first ${\cal N}$ excited
states of $V_{\cal N}^{-}$ have a finite convergence radius in $g^2$,
and if $w'(0)<0$, those of $V_{\cal N}^{+}$ have a finite convergence radius
in $g^2$. 
It is well-known that perturbative
expansions of usual quantum mechanics become divergent series \cite{GZ},     
thus this behavior means that all the possible divergent parts of the
perturbative corrections vanish in type A models. 

To prove the non-renormalization theorem, we adjust $c_1=1$ and
$c_2=0$ in Eq.(\ref{sec_e:hc}) and introduce $\eta(gq)$ as follows,
\begin{eqnarray}
h(q)=\frac{1}{g}\eta(gq),
\label{sec_e:nrh}
\end{eqnarray}
where 
\begin{eqnarray}
\eta(gq)=\int_0^{gq}dx_1e^{\int^{x_1}_0dx_2e(x_2)}.
\end{eqnarray}
Then we consider the characteristic equations of ${\bf S}^{\pm}$.
(See, Eqs.(\ref{sec_e:men}) and (\ref{sec_e:nen})). 
Since $\phi^{-}_n$ and $\phi^{+}_n$ behave as
\begin{eqnarray}
\phi^{-}_n(q)&=&U^{-1}(0)(q^{n-1}+O(g)) e^{-w'(0)q^2/2}, 
\nonumber\\
\phi^{+}_n(q)&=&U(0)V(0)(q^{n-1}+O(g)) e^{w'(0)q^2/2},  
\label{sec_e:phi+-}
\end{eqnarray}
either of the eigenstates of ${\bf S}^{-}$ or ${\bf S}^{+}$ are
normalizable,  at least in the perturbation theory.
Thus if $w'(0)>0$, the eigenvalues of ${\bf S}^{-}$ give exact
spectra of $H_{\cal N}^{-}$ in the perturbation theory, and if
$w'(0)<0$, the eigenvalues of ${\bf S}^{+}$ give those of $H_{\cal
N}^{+}$ in the perturbation theory.\footnote{If $\phi_n^{+}$ or
$\phi_n^{-}$ are normalizable without expanding by $g$, the spectra are
really exact. } 

Equation (\ref{sec_e:phi+-}) shows also that 
either linear combinations of $\phi_n^{+}$ or $\phi_n^{-}$ give the
first ${\cal N}$ eigenstates of the harmonic potentials
(\ref{sec_e:vlg}).
We especially notice here that if $w'(0)>0$, 
the first ${\cal N}$ eigenstates of $V_{\cal N}^{-}$ 
can be given by suitable linear combinations of $\phi^{-}_n$, and if
$w'(0)<0$, 
the first ${\cal N}$ eigenstates of $V_{\cal N}^{+}$
can be given by suitable linear combinations of $\phi^{+}_n$. 
Thus if $w'(0)>0$, 
all order of
the perturbative series for the first ${\cal N}$ excited energies of
$V_{\cal N}^{-}$ are given by the eigenvalues of ${\bf S}^{-}$,
and if $w'(0)<0$, those of $V_{\cal N}^{+}$ are given by ${\bf S}^{+}$.
In appendix \ref{appendixa}, we will show that the characteristic
equations of ${\bf S}^{\pm}$ are polynomials of $g^2$. 
Therefore, the eigenvalues of ${\bf S}^{\pm}$ have a finite
convergence radius in $g^2$. 
Thus the theorem is proved.

As far as we know, the Mother Hamiltonians of all known type A models are
polynomials of the original Hamiltonian, and the following relation
holds,
\begin{eqnarray}
{\cal H}_{\cal N}=\frac{1}{2}
{\rm det}{\bf M}^{-}_{\cal N}({\bf H}_{\cal N})
=\frac{1}{2}{\rm det}{\bf M}^{+}_{\cal N}({\bf H}_{\cal N})
\end{eqnarray}
where ${\bf M}^{\pm}_{\cal N}$ are given by
Eq.(\ref{sec_p:misnit}).
When ${\cal N}=2$, we can prove this generally.
We conjecture that this holds for arbitrary ${\cal N}$ in the type A models.

\section{${\cal N}$-fold supersymmetry in multi-dimensional quantum mechanics}
\label{nsims}

Finally, we will give a possible extension of ${\cal N}$-fold
supersymmetry in multi-dimensional quantum mechanics.
We denote the bosonic coordinates as $q_i$ $(i=1,\cdots, n_b)$ and the
fermionic ones as $\psi_i$ $(i=1,\cdots, n_f)$.
The fermionic coordinates satisfy
\begin{eqnarray}
\{\psi_i,\psi_j\}
=\{\psi_i^{\dagger},\psi_j^{\dagger}\}=0, 
\quad
\{\psi_i,\psi_j^{\dagger}\}=\delta_{i,j}. 
\end{eqnarray}
The Hamiltonian of the ${\cal N}$-fold supersymmetric system is
defined by 
\begin{eqnarray}
{\bf H}_{\cal N}=\sum_{i,j}H^{- (i,j)}_{\cal N}\psi_i\psi^{\dagger}_j
+\sum_{i,j}H^{+(i,j)}_{\cal N}\psi_i^{\dagger}\psi_j, 
\end{eqnarray}
where 
\begin{eqnarray}
H^{-(i,j)}_{\cal N}=\frac{1}{2}\sum_{k,l}G_{\cal N}^{-}{}_{k,l}^{(i,j)}p_kp_l 
+V_{\cal N}^{-}{}^{(i,j)},
\nonumber\\
H^{+(i,j)}_{\cal N}=\frac{1}{2}\sum_{k,l}G_{\cal N}^{+}{}_{k,l}^{(i,j)}p_kp_l 
+V_{\cal N}^{+}{}^{(i,j)}.
\end{eqnarray}
Here $G_{\cal N}^{\pm}{}^{(i,j)}_{k,l}$ and 
$V_{\cal N}^{\pm}{}^{(i,j)}$ are functions
of $q_i$ $(i=1,\cdots,n_b)$.
The ${\cal N}$-fold supercharges are generalized as 
\begin{eqnarray}
Q_{\cal N}=\sum_iP^{i \dagger}_{\cal N}\psi_i
,\quad 
Q_{\cal N}^{\dagger}=\sum_iP^{i}_{\cal N}\psi_i^{\dagger},
\end{eqnarray}
where $P^i_{\cal N}$ is an ${\cal N}$-th order polynomial of the
momenta $p_m\equiv -i\partial_m, (m=1,\cdots, n_b)$. 
To satisfy the following ${\cal N}$-fold superalgebra,
\begin{eqnarray}
\{Q_{\cal N}, Q_{\cal N}\}=\{Q_{\cal N}^{\dagger}, Q_{\cal N}^{\dagger}\}=0,
\quad
[Q_{\cal N}, {\bf H}_{\cal N}]=[Q^{\dagger}, {\bf H}_{\cal N}]=0,
\label{sec_nsims:sa}
\end{eqnarray}
we put the following conditions,
\begin{eqnarray}
[P_{\cal N}^i, P_{\cal N}^j]=0, 
\label{sec_nsims:sa1}
\end{eqnarray}
\begin{eqnarray}
H^{+(i,j)}_{\cal N}P_{\cal N}^k=P_{\cal N}^iH^{-(j,k)}_{\cal N}, 
\quad
H^{-(i,j)}_{\cal N}P_{\cal N}^k=H^{-(i,k)}_{\cal N}P_{\cal N}^j,
\quad
P_{\cal N}^iH^{+(j,k)}_{\cal N}=P_{\cal N}^jH^{+(i,k)}_{\cal N}.
\label{sec_nsims:sa2}
\end{eqnarray}
Equation (\ref{sec_nsims:sa1}) comes from the former equation in
(\ref{sec_nsims:sa}) and Eq.(\ref{sec_nsims:sa2}) comes from the latter.

When ${\cal N}=1$ and $n_b=n_f$, 
Eqs.(\ref{sec_nsims:sa1}) and (\ref{sec_nsims:sa2}) have the following
solution,
\begin{eqnarray}
P_1^{i}=p_i-i\partial_{i}h, \quad  
H_1^{-(i,j)}=P_1^{i\dagger}P_1^{j}, \quad
H_1^{+(i,j)}=P_1^{i}P_1^{j\dagger},
\end{eqnarray}
where $h$ is a function of $q_i (i=1,\cdots,n_b)$.
This reproduces ordinary supersymmetry in multi-dimensional quantum
mechanics in Ref.\cite{Wit2}.

Extensions of supersymmetry in multi-dimensional quantum mechanics
attempted in Refs.\cite{AIN,AIN3,KP2} correspond to ours with $n_f=1$ and
$n_b=2$.  
When $n_f=1$, Eq.(\ref{sec_nsims:sa2}) is simplified since the latter
two equations become trivial.

\vskip 1cm
\centerline{\Large \bf Acknowledgment}
\vskip 1cm
The authors would like to thank M. Ioffe, C. Dunning and  M. S. Plyushchay 
for correspondences.
H. Aoyama's work was supported in part by the Grant-in-Aid for
Scientific Research No.10640259.
T. Tanaka's work was supported in part by a JSPS research fellowship.

\vskip 1cm
\centerline{\Large\bf Appendix}
\vskip 1cm
\appendix
\section{$g$-dependence of ${\bf S}^{\pm}$ in type A models}   
\label{appendixa}
Here we will prove that if we introduce the coupling constant $g$ by
Eqs.(\ref{sec_e:defg}) and (\ref{sec_e:nrh}), 
the matrices ${\bf S}^{\pm}$ in
section.\ref{sec_e_typea} have the following forms,
\begin{eqnarray}
{\bf S}^{\pm}_{n,m}=g^{-n}{\cal P}^{\pm}_{n,m}(g^2)g^{m},
\label{sec_a:stp}
\end{eqnarray}
where ${\cal P}_{n,m}^{\pm}(g^2)$ are polynomials of $g^2$.
If Eq.(\ref{sec_a:stp}) holds, the characteristic equations becomes the
following polynomials of $g^2$:
\begin{eqnarray}
&&{\rm det}{\bf M}^{\pm}_{\cal N}(E_n)
={\rm det}({\cal P}^{\pm}(g^2)-E_n{\bf I})=0.
\end{eqnarray}

Eq.(\ref{sec_a:stp}) may be proved by induction, which we will prove
explicitly carry out for ${\bf S}^{-}$.
In a similar manner, Eq.(\ref{sec_a:stp}) for ${\bf S}^{+}$ can be
proved.  
First, we calculate ${\bf S}^{-}_{n,{\cal N}}$ by Eq.(\ref{sec_e:sncn}).
Since ${\bf S}^{-}_{n,{\cal N}}$ does not depend on $q$, 
it is evaluated by 
\begin{eqnarray}
{\bf S}^{-}_{n,{\cal N}}
=\left.
\frac{P_{{\cal N}-1}H^{-}_{\cal N}\phi^{-}_n}{P_{{\cal N}-1}\phi^{-}_{\cal N}}
\right|_{q=0} 
.
\end{eqnarray}
A straightforward calculation shows that
\begin{eqnarray}
H^{-}_{\cal N}\phi^{-}_{n}=g^{1-n}U^{-1}(q)
\left(
{\cal F}^0_{{\cal N},n}(gq)+g^2{\cal F}^1_{{\cal N},n}(gq)
\right), 
\end{eqnarray}
where 
\begin{eqnarray}
{\cal F}^0_{{\cal N},n}(gq)=&&
(n-1)w(gq)\eta'(gq)\eta(gq)^{n-2}
+\frac{1}{2}w'(gq)\eta(gq)^{n-1}
\nonumber\\
&&-\frac{{\cal N}-1}{2}e(gq)w(gq)\eta(gq)^{n-1}
-\frac{{\cal N}}{2}w'(gq)\eta(gq)^{n-1},
\nonumber\\
{\cal F}^1_{{\cal N},n}(gq)=&&
-\frac{n-1}{2}\eta''(gq)\eta(gq)^{n-2}
-\frac{(n-1)(n-2)}{2}\eta'(gq)^2\eta(gq)^{n-3}
\nonumber\\
&&+\frac{({\cal N}-1)(2{\cal N}-1)}{12}e(gq)^2\eta(gq)^{n-1}
-\frac{{\cal N}^2-1}{12}e'(gq)\eta(gq)^{n-1}
\nonumber\\
&&+\frac{{\cal N}({\cal N}-1)}{4}e'(gq)\eta(gq)^{n-1}.
\end{eqnarray}
We also obtain 
\begin{eqnarray}
P_{\cal M}&=&(-i)^{\cal M}U^{-1}(q)\tilde{P}_{\cal M}U(q)
\nonumber\\ 
&=&(-ig)^{\cal M}
U^{-1}(q)\left(\frac{d}{d(gq)}-({\cal M}-1)e(gq)\right)
\left(\frac{d}{d(gq)}-({\cal M}-2)e(gq)\right)
\nonumber\\
&&\times\cdots\times\left(\frac{d}{d(gq)}-e(gq)\right)\frac{d}{d(gq)}U(q)
\nonumber\\
&\equiv& (-ig)^{\cal M}U^{-1}(q)\prod^{{\cal M}-1}_{k=0}
\left(\frac{d}{d(gq)}-ke(gq)\right)U(q).
\end{eqnarray}
Using them, we obtain
\begin{eqnarray}
\left.P_{{\cal N}-1}H^{-}_{\cal N}\phi^{-}_n\right|_{q=0}
=&&(-i)^{{\cal N}-1}g^{{\cal N}}U^{-1}(0)
\prod_{k=0}^{{\cal N}-2}\left(\frac{d}{d(gq)}-ke(gq)\right)
\nonumber\\
&&\times
\left.\left({\cal F}^0_{{\cal N},n}(gq)+g^2{\cal F}^1_{{\cal N},n}(gq)
\right)\right|_{q=0}g^{-n}, 
\end{eqnarray}
and 
\begin{eqnarray}
\left.P_{{\cal N}-1}\phi^{-}_{\cal N}\right|_{q=0}=
(-i)^{{\cal N}-1}({\cal N}-1)!
U^{-1}(0),
\end{eqnarray}
where we have used $h'(0)=\eta'(0)=1$.
Thus ${\cal P}^{-}_{n,{\cal N}}(g^2)$ becomes
\begin{eqnarray}
{\cal P}^{-}_{n,{\cal N}}(g^2)
=\frac{1}{({\cal N}-1)!}\prod_{k=0}^{{\cal N}-2}
\left.\left(\frac{d}{d(gq)}-ke(gq)\right)
\left({\cal F}^0_{{\cal N},n}(gq)+g^2{\cal F}^1_{{\cal N},n}(gq)
\right)\right|_{q=0}
\end{eqnarray}

Next we assume that the matrices ${\bf S}^{-}_{n,{\cal N}-k}$ for
$k=0,\cdots,m$ have the forms ${\bf S}_{n,{\cal N}-k}=g^{-n}{\cal
P}^{-}_{n,{\cal N}-k}(g^2)g^{{\cal N}-k}$ and ${\cal P}^{-}_{n,{\cal
N}-k}(g^2)$ is a polynomial of $g^2$. 
Then ${\bf S}^{-}_{n,{\cal N}-m-1}$ is calculated by
\begin{eqnarray}
{\bf S}^{-}_{n,{\cal N}-m-1}&=&\left.
\frac{P_{{\cal N}-m-2}(H^{-}_{\cal N}\phi^{-}_n
-\sum_{k=0}^{m}{\bf S}^{-}_{n,{\cal N}-k}\phi^{-}_{{\cal N}-k})}
{P_{{\cal N}-m-2}\phi^{-}_{{\cal N}-m-1}}\right|_{q=0} 
\nonumber\\
&=&\left.
\frac{P_{{\cal N}-m-2}H^{-}_{\cal N}\phi^{-}_n}
{P_{{\cal N}-m-2}\phi^{-}_{{\cal N}-m-1}}
\right|_{q=0}
-\sum_{k=0}^{m}{\cal P}^{-}_{n,{\cal N}-k}(g^2)g^{{\cal N}-k-n}
\left.
\frac{P_{{\cal N}-m-2}\phi^{-}_{{\cal N}-k}}
{P_{{\cal N}-m-2}\phi^{-}_{{\cal N}-m-1}} 
\right|_{q=0}.
\end{eqnarray}
From this, we find that ${\bf S}_{n,{\cal N}-m-1}$ also has the form
${\bf S}_{n,{\cal N}-m-1}=g^{-n}{\cal P}^{-}_{n,{\cal N}-m-1}(g^2)g^{{\cal
N}-m-1}$ and ${\cal P}_{n,{\cal N}-m-1}^{-}(g^2)$ becomes
\begin{eqnarray}
&&{\cal P}^{-}_{n,{\cal N}-m-1}(g^2)
\nonumber\\
&&=\frac{1}{({\cal N}-m-2)!}
\prod_{k=0}^{{\cal N}-m-3}
\left.\left(\frac{d}{d(gq)}-ke(gq)\right)
\left({\cal F}^0_{{\cal N},n}(gq)+g^2{\cal F}^1_{{\cal N},n}(gq)
\right)\right|_{q=0}
\nonumber\\
&&-\sum_{k=0}^{m}
\frac{{\cal P}^{-}_{n,{\cal N}-k}(g^2)}{({\cal N}-m-2)!}
\left.\prod_{k=0}^{{\cal N}-m-3}
\left(\frac{d}{d(gq)}-ke(gq)\right)
\eta^{{\cal N}-k-1}(gq)
\right|_{q=0}.
\end{eqnarray}
This is a polynomial of $g^2$.


\begin{thebibliography}{99}
\def\J#1#2#3#4{{\sl #1} {\bf #2} (#3) #4}
\def\PL{Phys. Lett.}
\def\NP{Nucl. Phys.}

\bibitem{SW}N. Seiberg and E. Witten, 
\J{\NP}{B431}{1994}{484}.



\bibitem{AKOSW} H. Aoyama, H. Kikuchi, I. Okouchi, M. Sato and
  S. Wada, \J{\PL}{B424}{1998}{93}.



\bibitem{AKOSW2}
H. Aoyama, H. Kikuchi, I. Okouchi, M. Sato and S. Wada, 
\J{\NP}{B553}{1999}{644}.






\bibitem{RR}
D. J. Rowe and A. Ryman, \J{J. Math. Phys.}{23}{1982}{732}.
\bibitem{BY}
I. I. Balitsky and A. V. Yung, \J{\PL}{B168}{1986}{13}.
\bibitem{Sil}
P. G. Silvetrov, \J{Sov. J. Nucl. Phys.}{51}{1990}{1121}.
\bibitem{AK}
H. Aoyama and H. Kikuchi, \J{\NP}{B369}{1992}{219}.
\bibitem{AW}
H. Aoyama and S. Wada, \J{\PL}{B349}{1995}{279}.
\bibitem{HS} T. Harano and M. Sato, {\sl hep-ph}/9703457. 
\bibitem{AKHSW} 
H. Aoyama, H. Kikuchi, T. Harano, M. Sato and S. Wada, 
\J{Phys. Rev. Lett.}{79}{1997}{4052}.
\bibitem{AKHOSW}H. Aoyama, H. Kikuchi, T. Harano, I. Okouchi, M. Sato 
and S. Wada,
\J{Prog. Theor. Phys. Supplement}{127}{1997}{1}.

\bibitem{Bog}E. B. Bogomolny, \J{\PL}{B91}{1980}{431}.

\bibitem{Wit}
E. Witten,
\J{\NP}{B188}{1981}{513}.
\bibitem{Wit2}
E. Witten,
\J{\NP}{B202}{1982}{253}.




\bibitem{AIS}
A. A. Andrianov, M. V. Ioffe and V. P. Spiridonov,
\J{\PL}{A174}{1993}{273}.
\bibitem{AICD}
A. A. Andrianov, M. V. Ioffe, F.Cannata and J.-P.Dedonder,
\J{Int. J. Mod. Phys.}{A10}{1995}{2683}.
\bibitem{AIN}
A. A. Andrianov, M. V. Ioffe and D. N. Nishnianidze,
\J{\PL}{A201}{1995}{103}.
\bibitem{AIN2}
A. A. Andrianov, M. V. Ioffe and D. N. Nishnianidze,
\J{Theor. Math. Phys.}{104}{1995}1129.

\bibitem{BS}
V. G. Bagrov and B. F. Samsonov,
\J{Theor. Math. Phys.}{104}{1995}1051.
\bibitem{Sam}
B. F. Samsonov,
\J{Mod. Phys. Lett.}{A11}{1996}1563.
\bibitem{BS2}
V. G. Bagrov and B. F. Samsonov,
\J{Phys. Part. Nucl.}{28}{1997}{374}.
\bibitem{Sam2}
B. F. Samsonov,
\J{\PL}{A263}{1999}{274}.


\bibitem{AST}H. Aoyama, M. Sato and T. Tanaka, 
\J{\PL}{B503}{2001}{423}.

\bibitem{ASTY}
H. Aoyama, M. Sato, T. Tanaka and M. Yamamoto,
\J{\PL}{B498}{2001}{117}.


\bibitem{Ply}
M. Plyushchay,
\J{Int. J. Mod. Phys.}{A15}{2000}{3679}.
\bibitem{KP}
S. Klishevich and M. Plyushchay,
\J{Mod. Phys. Lett.}{A14}{1999}{2739}.

\bibitem{Ros1}
J. O. Rosas-Ortiz,
\J{J. Phys.}{A31}{1998}{10163}.
\bibitem{KS}
A. Khare and U. Sukhatme,
\J{J. Math. Phys.}{40}{1999}{5473}.
\bibitem{FH}
D. J. Fern\'andez C. and V. Hussin,
\J{J. Phys.}{A32}{1999}{3630}.
\bibitem{FNN}
D. J. Fern\'andez C., J. Negro and L. M. Nieto,
\J{\PL}{A275}{2000}{338}.

\bibitem{Tur}
A. V. Turbiner,
\J{Commun. Math. Phys.}{118}{1988}{467}

\bibitem{Shi}
M. A. Shifman, 
\J{Int. J. Mod. Phys.}{A12}{1989}{2897}.

\bibitem{Ush}
A. G. Ushveridze, {\sl Quasi-Exactly Solvable Models in Quantum
	Mechanics}, 
(IOP Publishing, Bristol, 1994),
and references cited therein.

\bibitem{KP1}
S. M. Klishevich and M. S. Plyushchay,
{\sl hep-th}/0012023.


\bibitem{DDT}P. Dorey, C. Dunning and R. Tateo,
{\sl hep-th}/0103051.

\bibitem{GZ}For a review, see {\sl Large-Order Behavior of Perturbation
	Theory}, ed. J.C. Le Guillou and J. Zinn-Justin (North-Holland,
	Amsterdam, 1990).
 
\bibitem{AIN3}
A. A. Andrianov, M. V. Ioffe and D. N. Nishnianidze,
{\sl solv-int}/9605007


\bibitem{KP2}
S. M. Klishevich and M. S. Plyushchay,
{\sl hep-th}/0105135.


 
\end{thebibliography}
\end{document}